\documentclass[fleqn,usenatbib]{mnras}

\usepackage{newtxtext,newtxmath}

\usepackage[T1]{fontenc}

\DeclareRobustCommand{\VAN}[3]{#2}
\let\VANthebibliography\thebibliography
\def\thebibliography{\DeclareRobustCommand{\VAN}[3]{##3}\VANthebibliography}

\usepackage{graphicx}	
\usepackage{amsmath}	
\def\sw {Swift J0549.7-6812}
\def\m2002 {[M2002] SMC 12102}
\newcommand{\bexrb}{BeXRB}
\newcommand{\bexrbs}{BeXRBs}

\title[ \sw] {A rare outburst from the stealthy \bexrb ~system \sw. }

\author[M. J. Coe et al.]{M.~J. Coe$^{1}$\thanks{E-mail: mjcoe@soton.ac.uk},
  J.~A. Kennea$^{2}$, I. M. Monageng$^{3,4}$, D.A.H. Buckley $^{3}$, A. Udalski$^{5}$ \& P.~A. Evans$^{6}$\\
$^{1}$Physics \& Astronomy, The University of Southampton, SO17 1BJ, UK\\
$^{2}$Department of Astronomy and Astrophysics, The Pennsylvania State University, 525 Davey Lab, University Park, PA 16802, USA\\
$^{3}$ South African Astronomical Observatory, P.O Box 9, Observatory, 7935, Cape Town, South Africa\\
$^{4}$ Department of Astronomy, University of Cape Town, Private Bag X3, Rondebosch 7701, South Africa\\
$^{5}$Astronomical Observatory, University of Warsaw, Al. Ujazdowskie 4, 00-478 Warszawa, Poland\\
$^{6}$University of Leicester, Astrophysics Group, School of Physics \& Astronomy, University Road, Leicester LE1 7RH, UK\\
}

\date{Accepted XXX. Received YYY; in original form ZZZ}

\pubyear{2021}

\begin{document}
\label{firstpage}
\pagerange{\pageref{firstpage}--\pageref{lastpage}}
\maketitle

\begin{abstract}
\sw ~is an Be/X-ray binary system (\bexrb) in the Large Magellanic Cloud (LMC) exhbiting a $\sim$6s pulse period. Like many such systems the variable X-ray emission is believed to be driven by the underlying behaviour of the mass donor Be star. In this paper we report on X-ray observations of the brightest known outburst from this system which reached a luminosity of ${\sim8} \times 10^{37}$~erg$\cdot {\rm s}^{-1}$. These observations are supported by contemporaneous optical photometric observations, the first reported optical spectrum, as well as several years of historical data from OGLE and GAIA. The latter strongly suggest a binary period of 46.1d. All the observational data indicate that \sw ~is a system that spends the vast majority of its time in X-ray quiescence, or even switched off completely. This suggests that occasional observations may easily miss it, and many similar systems, and thereby underestimate the massive star evolution numbers for the LMC. 
\end{abstract}




\section{ Introduction}

\color{black}

The stellar systems that are known as \bexrbs\ are a large sub-group of High Mass X-ray Binaries (HMXB) population that are characterised by consisting of a massive mass donor star, normally an OBe type, and an accreting compact object, normally a neutron star. The Small Magellanic Cloud (SMC) has been established for quite a while now as containing the largest known collection of \bexrbs\ ~\citep{ck2015, hs2016}. However recent more comprehensive studies of the Large Magellanic Cloud (LMC) are now rapidly increasing the numbers in that galaxy \citep{2023haberl}. The complex interactions between the two stars continues to produce unexpected surprises and the challenges are in the detailed understanding of the manner in which the two stars interact. Specifically, the often unpredictable behaviour of the mass donor OB-type star is believed to be major driver in the observed characteristics of such systems. However, if the Be star is relatively quiescent then long periods may elapse between such systems being detected as an X-ray source. The results presented here suggest that \sw ~falls into that category.

\color{black}

The 2013 outburst of \sw ~was serendipitously detected by the Swift X-ray Telescope (XRT; \citealt{burrows05}) as part of a set of observations of the LMC by \cite{2013krimma, 2013krimmb}. Those authors reported that the source was positionally consistent with a faint ROSAT source, 1RXS J055007.0-681451, but was not detected in any Chandra archival observations. Furthermore, they associated the X-ray source with a spectrally unclassified optical source, 2MASS 05500646-6814559. In addition, the X-ray data revealed a clear period of 6.2s which the authors interpreted as the probable spin period of a neutron star partner in a \bexrb system. Such a type of X-ray binary system was further supported by the observed hard X-ray spectrum.

Reported here are observations from Swift covering the full duration of the X-ray outburst phase. In addition I-band data are used to show the optical behaviour of the binary counterpart over several years. From these optical data it is seen that the pattern of changes in this system indicate very little activity over many years, consistent with the rarely known X-ray detections. However, there is a small but significant underlying regular modulation of the I-band measurements at a period of 46.1d which is interpreted here as the binary period of the system. 

The first optical spectrum of the counterpart is also presented here, and shows it to be a classic B0-0.5 IV-Ve star with a definite circumstellar disc. However, in this system it is probable that the neutron star orbital size is larger than the disc size almost all of the time. That could explain the rarity of any X-ray outbursts from \sw. Therefore the presence of this system, and its behaviour, may suggest the existence of many other \bexrb in the Magellanic Clouds that spend most of their time in a stealthy or quiescent mode, and may not yet be identified. This could have implications for the assessmnet of massive star formation rates in these galaxies.

\section{Observations}

\subsection{ X-rays - Swift}

\begin{figure}

	\includegraphics[width=8cm,angle=-00]{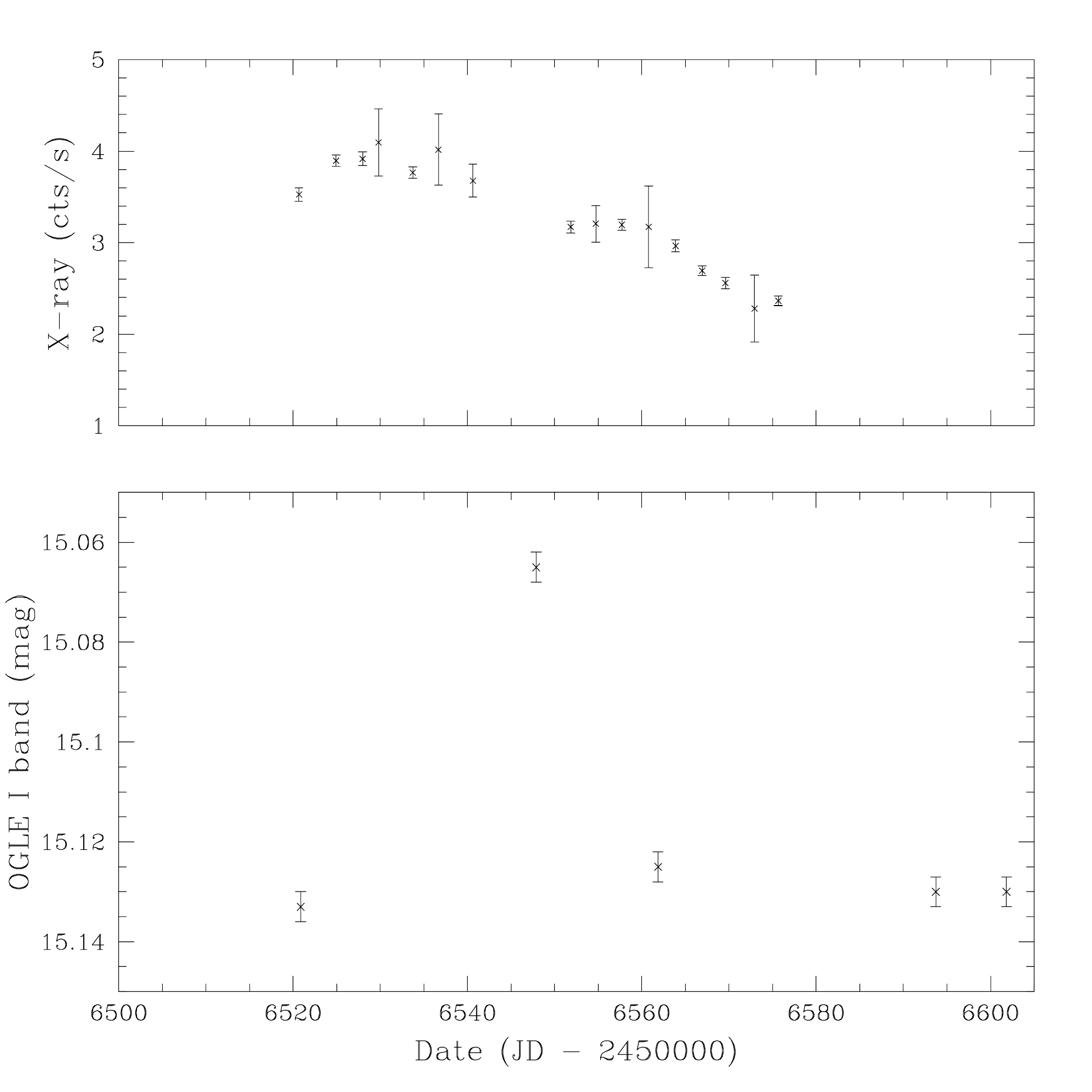}
    \caption{Swift XRT data (top panel). OGLE I-band data for the same time period (lower panel).   }
    \label{fig:swift_lc}
\end{figure}

\sw ~was detected by the Swift X-ray Telescope (XRT; \citealt{burrows05}) in August 2013 and the initial results reported in detail by \cite{2013krimma, 2013krimmb}.  Subsequently, a series of Window Timing observations were carried out which monitored the system till it was no longer detectable by the XRT. A detailed plot of the intensity of the outburst is shown in Figure ~\ref{fig:swift_lc} which shows that, at its peak, the XRT count rate was $ 3.9 \pm 0.1$ cts/s. 
 Using a standard LMC distance of 49.97~kpc \citep{2013Piet} and correcting for absorption fixed at the value derived from \cite{Willingale2013}, this corresponds to a peak 0.3-10~keV luminosity of ${(8.10 \pm 0.05)} \times 10^{37}$~erg$\cdot {\rm s}^{-1}$. 

The measured period history throughout the outburst is shown in Figure ~\ref{fig:swift_period}. As will be seen below, the probable binary period of this system is $\sim$46d and the Swift data set only cover a time frame of $\sim$50d. So it is not possible to disentangle any binary-driven period changes from accretion-driven ones over such a relatively short length of time.

The spectra are well-fitted with an absorbed power law spectrum throughout the outburst. The overall average fit gives a power law index of (0.88 $\pm$ 0.08) and an absorption of $N_{H}=(1.5 \pm 0.4) \times 10^{21} cm^{-2}$. These average values are driven by the brighter, earlier part of the outburst. However, there is some evidence of gradual change during the outburst, a possible increase in absorption and maybe a small drift in the power law index - see Figure ~\ref{fig:2panels}. Since the X-ray outburst is almost certainly driven by the passage of the neutron star through the out flowing material from the Be star in the form of a circumstellar disc, these small changes probably simply reflect the changing local density of the disc.

\begin{figure}

	\includegraphics[width=8cm,angle=-00]{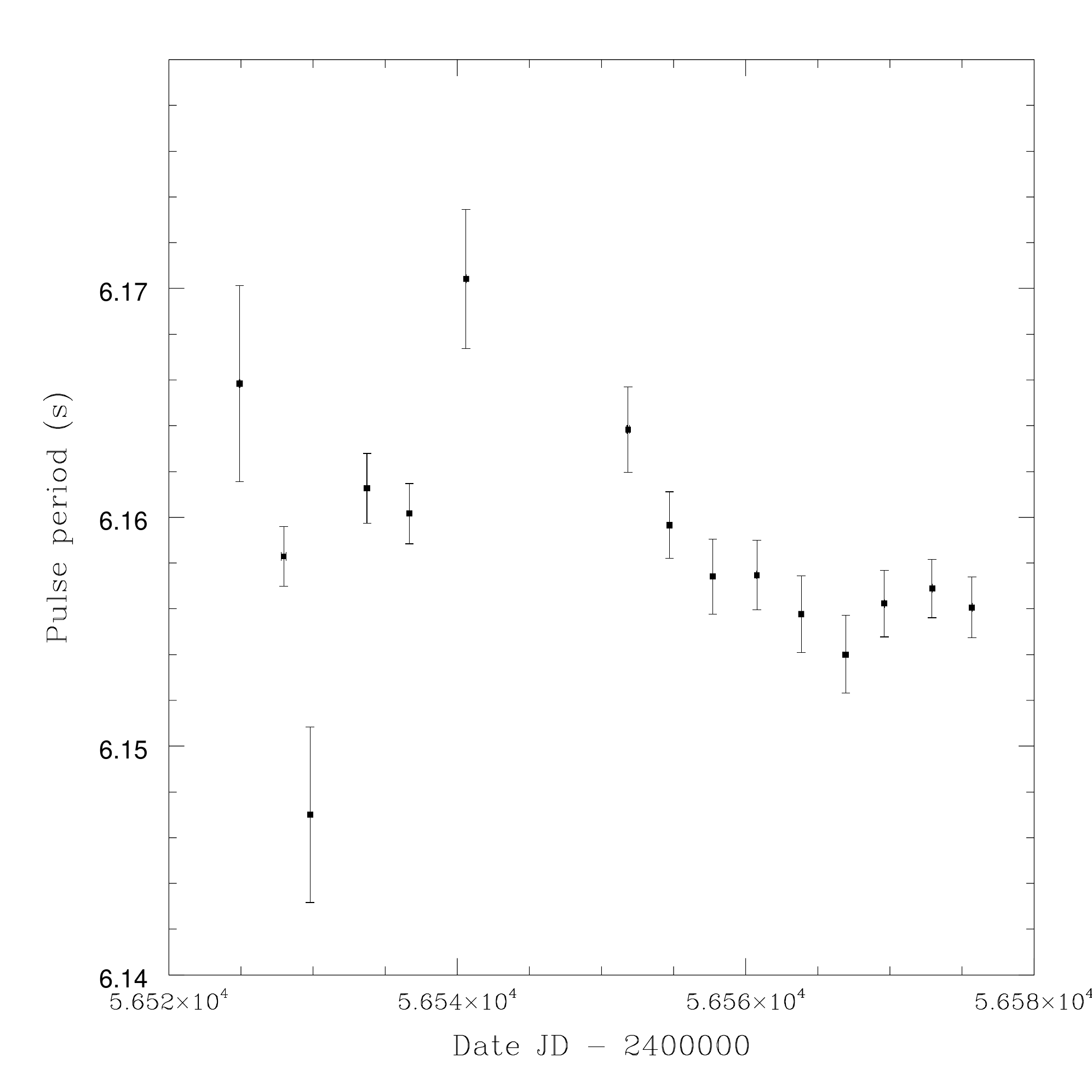}
    \caption{Swift XRT data showing measured periods throughout the outburst.}
    \label{fig:swift_period}
\end{figure}

\begin{figure}

	\includegraphics[width=8cm,angle=-00]{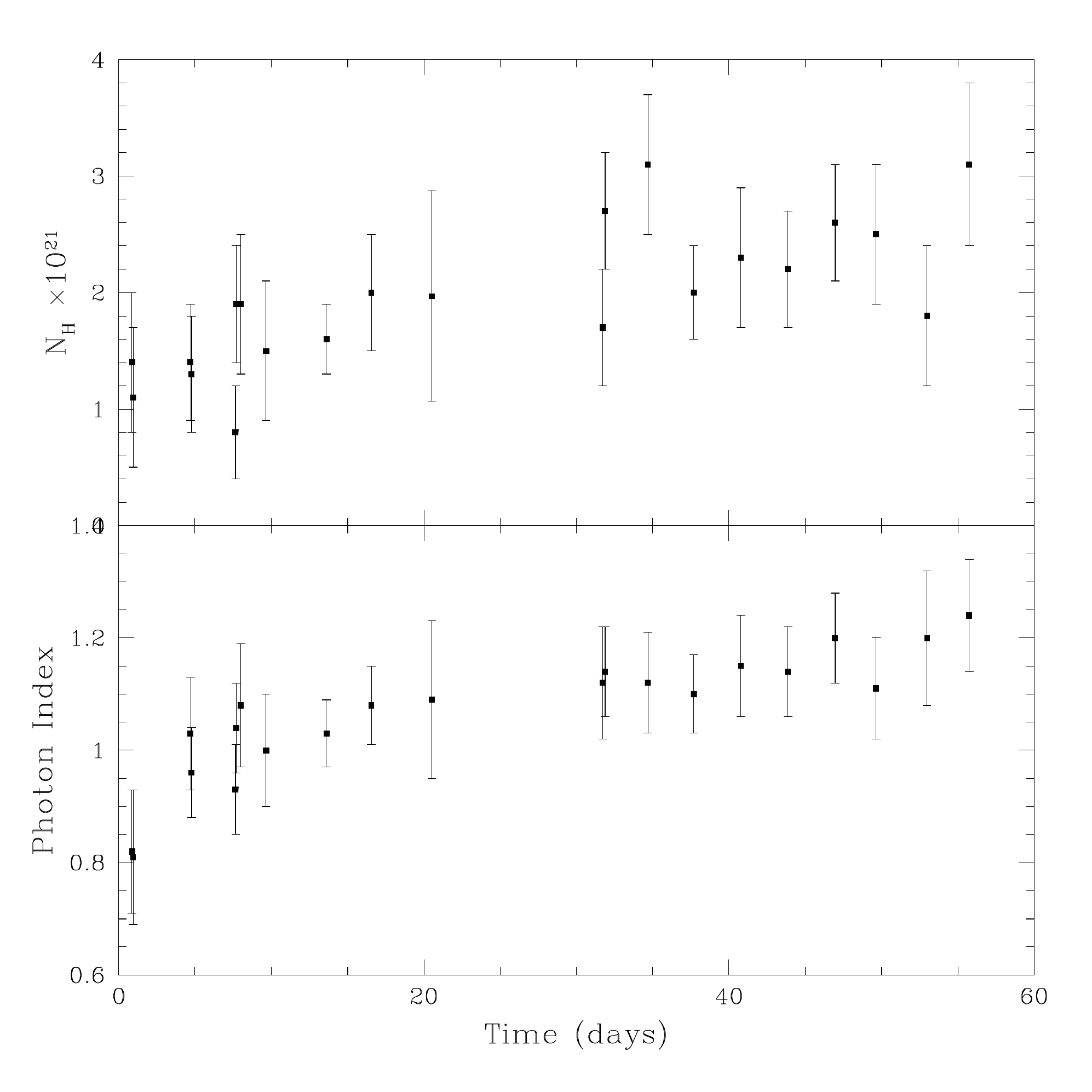}
    \caption{Swift XRT data showing measured spectral parameters throughout the outburst. The time axis starts at TJD 6524.0.}
    \label{fig:2panels}
\end{figure}



Long term observations over 10 years by the Burst Alert Telescope (BAT) \citep{2013krimmc}  on Swift have not detected any further X-ray outbursts since the one reported here in August 2013. This confirms the generally X-ray quiescent nature of this particular \bexrb ~system.

\subsection{OGLE IV}

The OGLE project \citep{Udalski2015} provides long term I-band photometry with a cadence of 1-3 days. The optical counterpart to \sw ~fell close to a gap between chips and hence was observed sporadically for $\sim$2000 days on two adjacent chips. It is identified in the OGLE catalogue as:\\
\\
OGLE IV (I band): LMC554.08.234D and LMC554.09.806D\\

\begin{figure}

	\includegraphics[width=8cm,angle=-00]{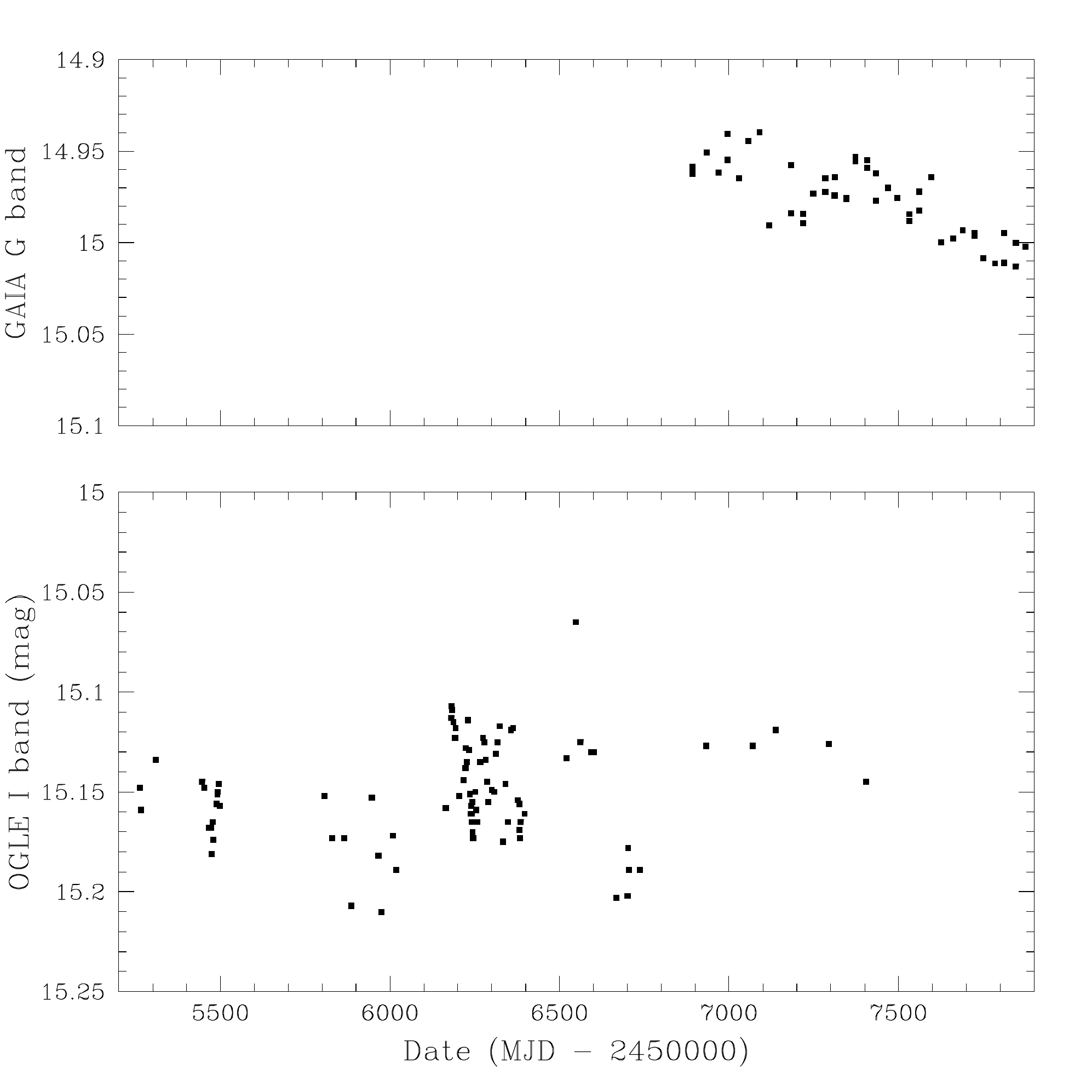}
    \caption{GAIA G band (top panel). OGLE I-band (lower panel). The brightest OGLE point was measured on TJD 6547.9 in the middle of the X-ray outburst (see also Figure ~\ref{fig:swift_lc}).  }
    \label{fig:ogle_lc}
\end{figure}

The I-band lightcurve produced from the OGLE IV observations that were clear of the chip gap are shown in Figure ~\ref{fig:ogle_lc}. The detailed lightcurve corresponding to the time of the X-ray outburst is shown in Figure ~\ref{fig:swift_lc}. It is noteworthy that there is one OGLE measurement that is brighter than all the others (I= 15.07) midway through the outburst at TJD 6547.9. By the time of the next OGLE measurement (TJD 6561.9) the source seems to have returned to the contemporaneous baseline of I=15.13. The data point just before the bright point was at TJD 6520.9, thereby constraining the maximum duration of the optical outburst to be 41d, though it could have been much shorter. This represents a very brief outburst time for a Be star.

\begin{figure}

	\includegraphics[width=8cm,angle=-00]{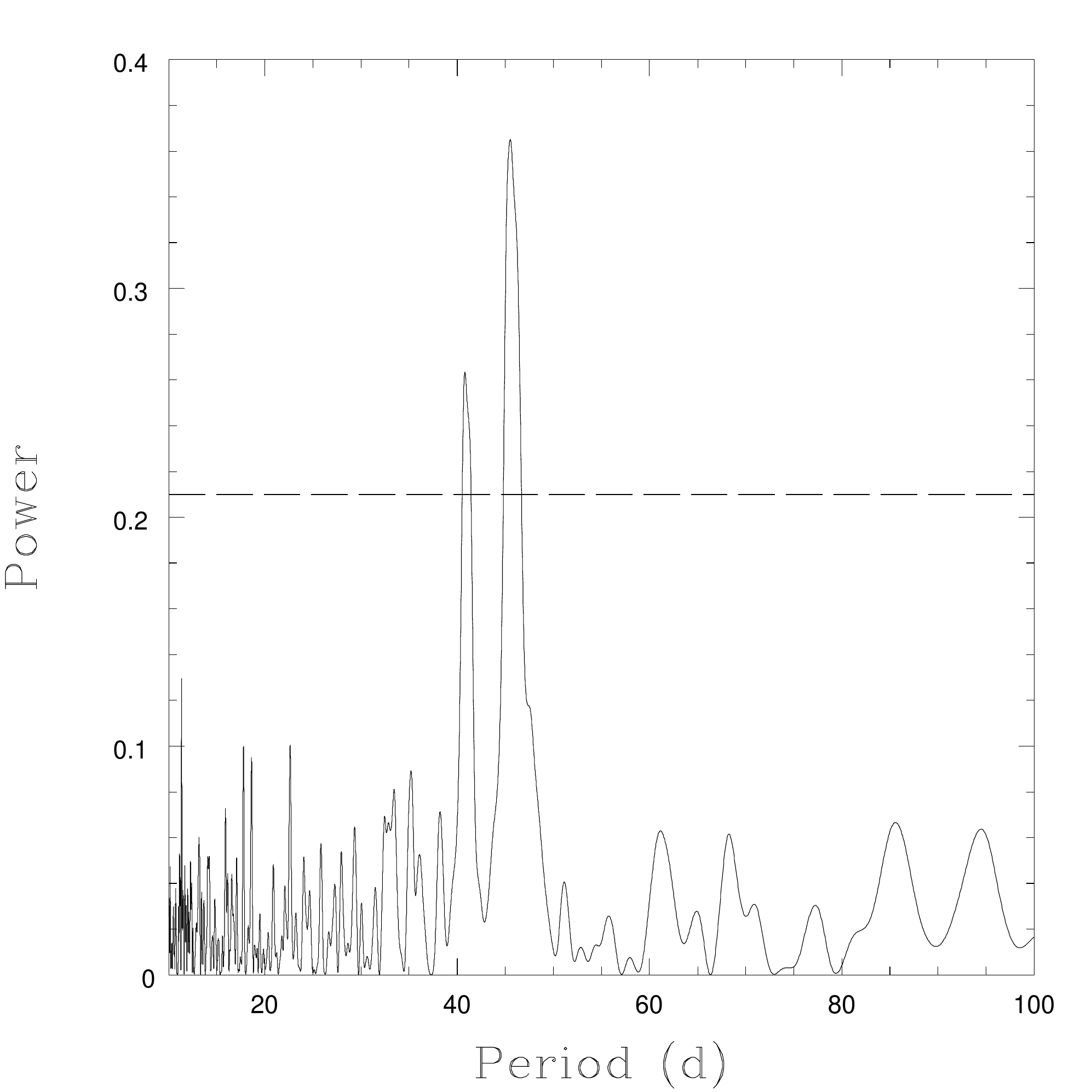}
    \caption{OGLE I-band power spectrum for the search range 10 -- 100d. The dashed line indicates the 1\% False Alarm Probability. The highest peak is at a period of 46.1d. }
    \label{fig:ogle_ps}
\end{figure}

These OGLE data were detrended and then searched for any significant periodicities using a Lomb-Scargle technique. A clear peak at $46.1\pm0.04$d was found - see Figure ~\ref{fig:ogle_ps}. A second peak representing the beating with the annual sampling is also visible.

\begin{figure}

	\includegraphics[width=8cm,angle=-00]{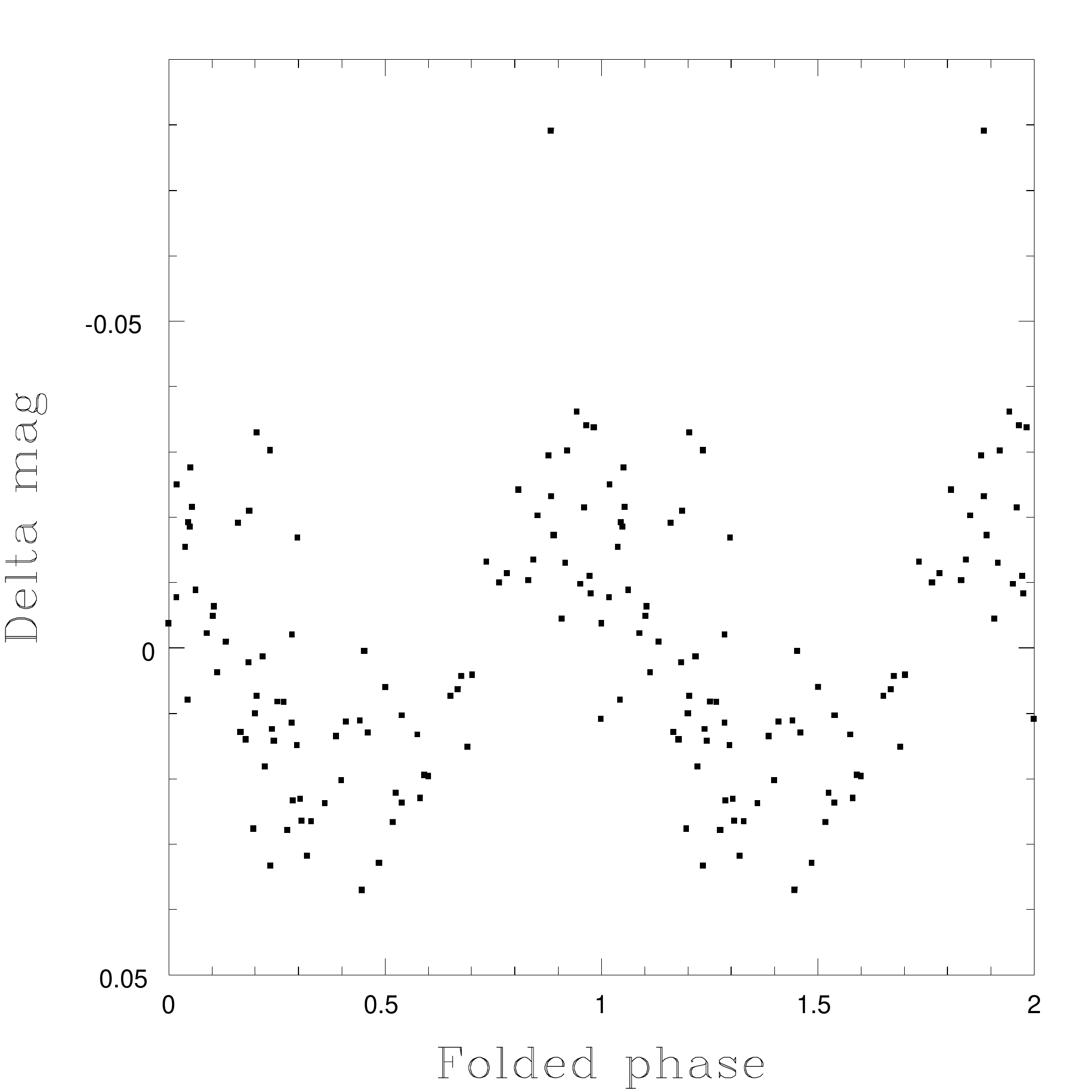}
    \caption{Detrended OGLE I-band folded at a period of 46.1d. Note the one high data point at a phase around 0.9 in this plot was recorded at the time of the X-ray outburst. }
    \label{fig:ogle_fold}
\end{figure}

Folding these OGLE data at this period shows a clear sinusoidal-like modulation with an amplitude of $\sim$0.05 magnitudes - see Figure ~\ref{fig:ogle_fold}. It is worth noting that the one bright data point corresponding to TJD 6547.8 which occurred at the centre of the X-ray outburst stands out clearly (close to phase 0.0 in the plot) from all the other OGLE measurements.

\subsection{ GAIA}

 The GAIA DR3 data release contains photometric time series data for \sw ~in three bandpasses - G, $G_{RP}$ and $G_{BP}$ \citep{2022gaia}. The data cover the period 22 August 2014 (TJD 56892) to 1 May 2017 (TJD 57874), so begin after the X-ray outburst. See Figure ~\ref{fig:ogle_lc}.
 
\begin{figure}

	\includegraphics[width=8cm,angle=-00]{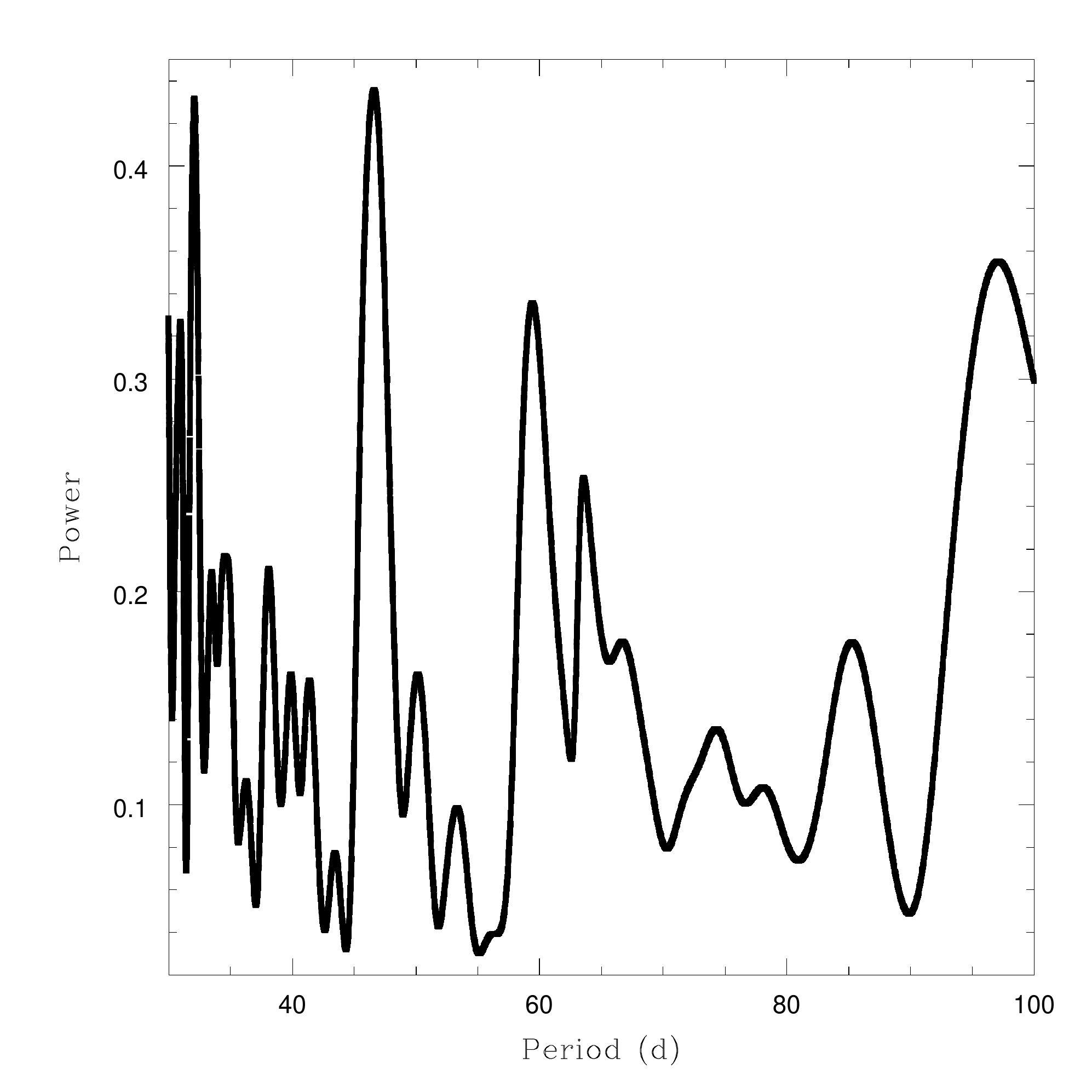}
    \caption{GAIA power spectrum for search range 30 -- 100d. Strongest peak is at 46.5d.}
    \label{fig:gaia_ps}
\end{figure}

 These data were searched using Lomb-Scargle techniques and though no independently 
 significant peak was detected, the highest peak in the power spectrum occurred at a period of 46.5d - see Figure ~\ref{fig:gaia_ps}. This strongly suggests the same periodic modulation is seen in the GAIA data that is seen in the OGLE data at 46.1d. But since the amplitude of the OGLE modulation is small and the GAIA time coverage much less, the period does not present itself as a significant result in the GAIA data alone.

\begin{figure}

	\includegraphics[width=8cm,angle=-00]{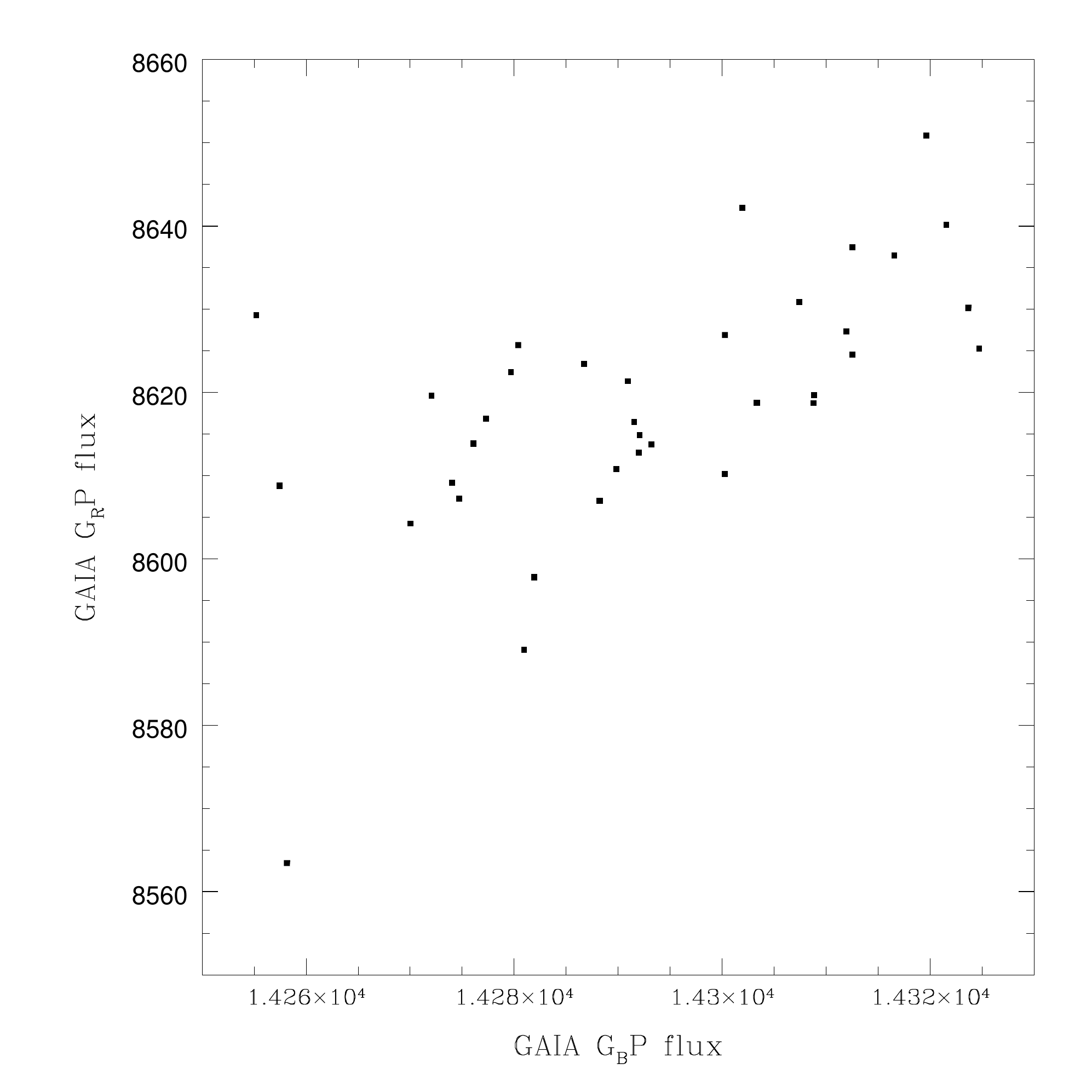}
    \caption{Plot showing observed colour between the GAIA filters $G_{RP}$ and $G_{BP}$ versus the overall brightness seen in the broad $G$ filter.}
    \label{fig:gaia_col}
\end{figure}

 Since the GAIA data are recorded in separate bandpass filters it is possible to look for colour variations seen between the $G_{RP}$ (red) and $G_{BP}$ (blue) filters compared with overall intensity changes seen in the $G$ filter - see Figure ~\ref{fig:gaia_col}. From this figure it can be seen that over the limited brightness range observed by GAIA there is essentially no colour change. This is perhaps not surprising if the Be star in \sw ~is showing little change in behaviour, as is evidenced also by the longer term OGLE observations.

\subsection{SALT}

\begin{figure*}
	\includegraphics[width=16cm,angle=-00]{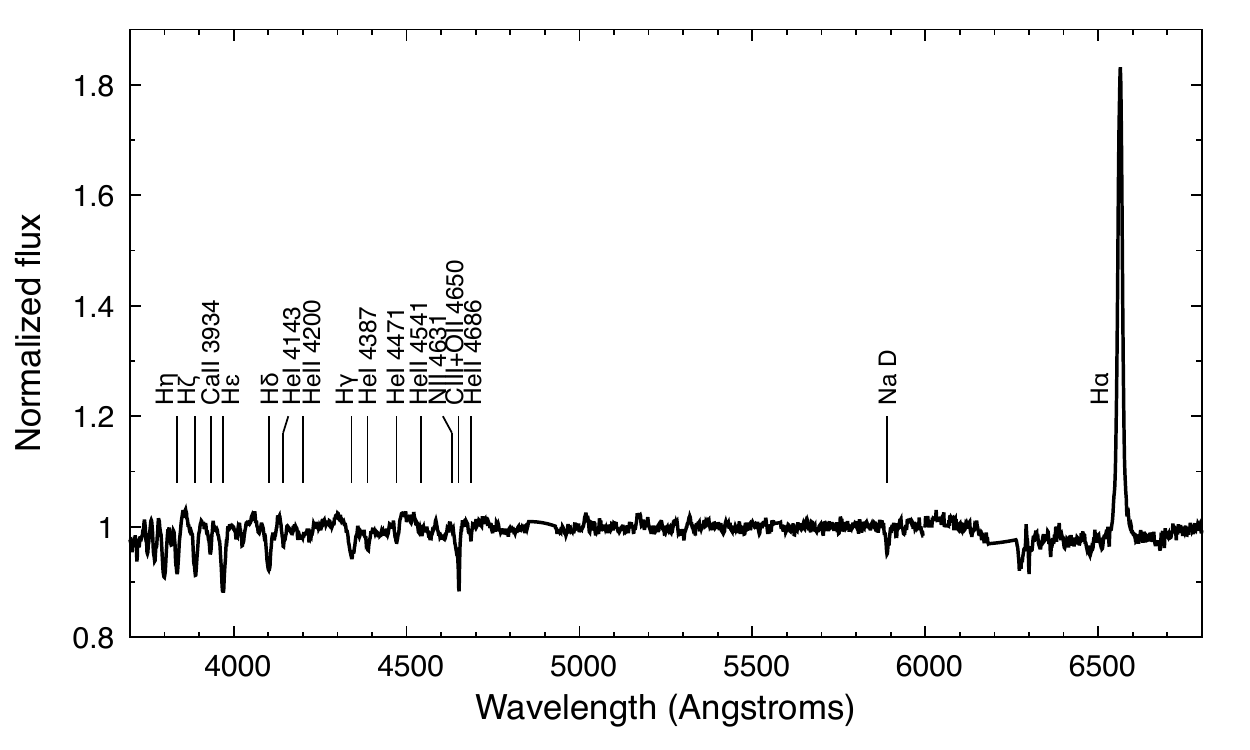}
    \caption{The normalized SALT RSS spectrum of \sw~with the different line species labeled at their expected rest wavelengths.}
    \label{fig:salt}
\end{figure*}

\begin{figure}
	\includegraphics[width=8cm,angle=-00]{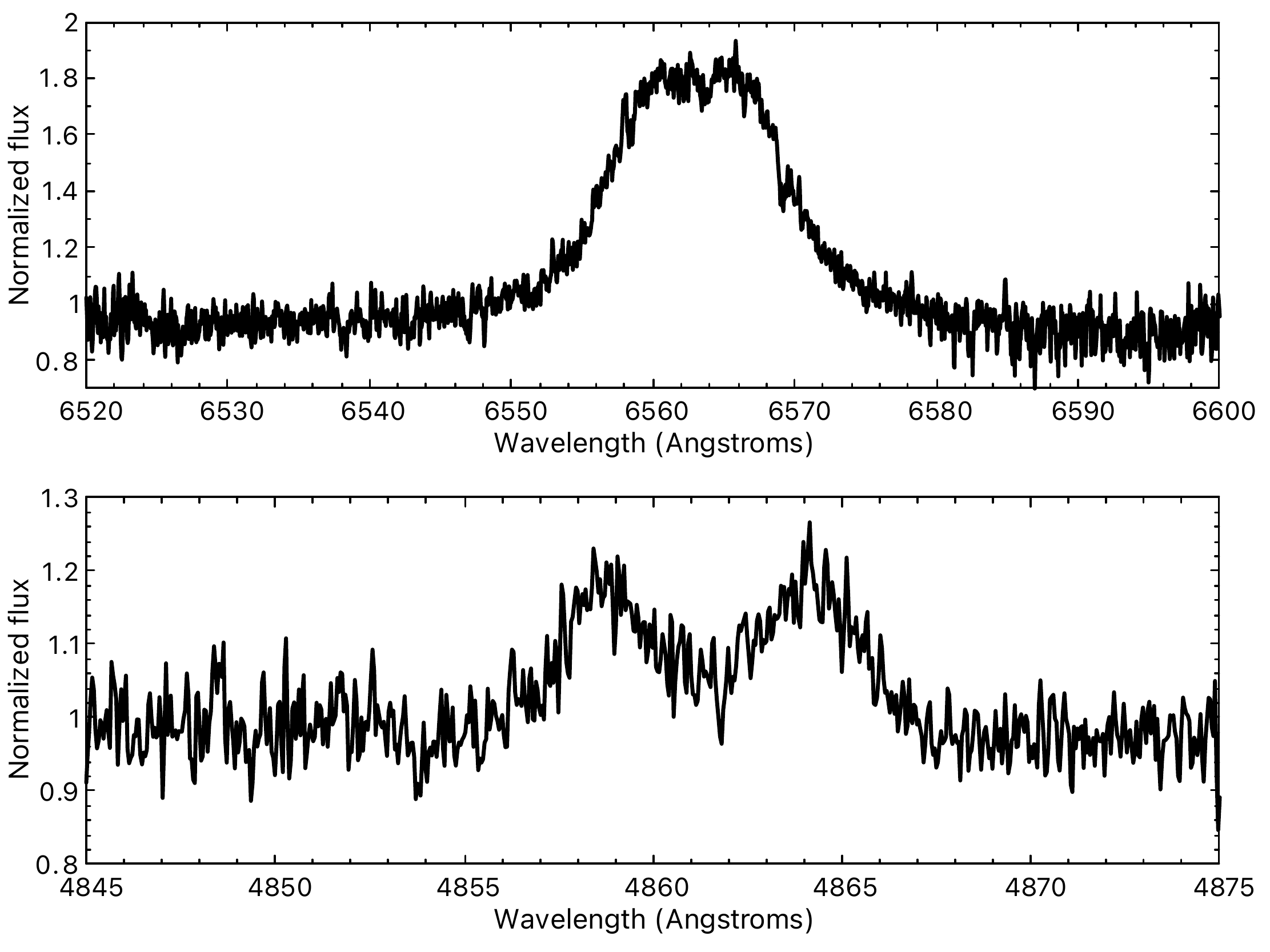}
    \caption{The H$\alpha$ and H$\beta$ emission lines from the SALT HRS observation}
    \label{fig:HRS_Ha}
\end{figure}

\sw ~was observed with the Southern African Large Telescope (SALT; \citealt{2006SPIE.6267E..0ZB}) using the Robert Stobie Spectrograph (RSS;\citealt{2003SPIE.4841.1463B,2003SPIE.4841.1634K}) on 25-02-2023 (MJD60000.81) and the High Resolution Spectrograph (HRS; \citealt{Bramall2010,Bramall2012,Crause2014}). The PG0700 grating was used for the RSS observation with a tilt angle of 4.6$^\circ$ and an exposure time of 900~seconds covering a wavelength range $3600 - 7400$~\AA. The SALT science pipeline \citep{2012ascl.soft07010C} was used to perform primary reductions, which include overscan corrections, bias subtraction, gain and amplifier cross-talk corrections. The remaining steps for the RSS observation, which comprised wavelength calibration, background subtraction and extraction of the one-dimensional spectrum were performed with \textsc{iraf}. The HRS observation was performed on 21-03-2023 (MJD60024.81) using the low-resolution mode and exposure time of 2400~seconds. For the HRS observation, the rest of the reduction steps which include background
subtraction, identification of arc lines, blaze function removal, and
merging of the orders were performed with the \textsc{midas feros} \citep{Stahl1999} and \textsc{echelle} \citep{Ballester1992}. The full details of the reduction procedure are described in \cite{Kniazev2016}.\\
Figure~\ref{fig:salt} shows the SALT RSS spectrum with the different line species labeled at their expected rest wavelengths. The H$\alpha$ emission line is shown in Figure~\ref{fig:HRS_Ha}. The spectra are corrected for the heliocenter and redshift of the LMC. The H$\alpha$ line is seen in strong emission, which is typical of Be stars due to the presence of the circumstellar disc. The  measured equivalent width (EW) of the H$\alpha$ emission line is -16.83 $\pm$ 0.49~\AA~ and -16.73 $\pm$ 0.49\AA~ for the RSS and HRS observations, respectively.

The blue spectral data may be used to spectrally classify the mass donor star. Based on the criteria in \cite{Evans2004} and referring to Figure~\ref{fig:salt}, the presence of the CIII and OII blend at 4650 A makes the spectral type earlier than B3; and the presence of He II 4686, together with weak/absence of HeII 4541 and HeII 4200 constrains the star to B0--0.5. 
Using the distance modulus of the LMC of 18.52 \citep{McConnachie2012} and the V-band mag of the target of 15.4, this results in an absolute magnitude of -3.1, which implies a luminosity class of IV-V. In summary, the spectral class of the massive companion is proposed to be B0-0.5 IV-Ve.

\section{Discussion}

\subsection {Comparing X-ray with optical lightcurves}

It is noteworthy that the one brightest OGLE point occurred at the time of peak optical brightness seen in the folded OGLE lightcurve (see Figure  ~\ref{fig:ogle_fold}). And  that it also coincided with the middle of the X-ray outburst (see Figure ~\ref{fig:swift_lc}). So using that as a reference point the ephemeris for the time of the optical peak brightness is given by Equation \ref{eq:1}: 

\begin{equation}
T_{opt} = 2456547.9 + N(46.09\pm{0.04}) ~\textrm{~JD}
\label{eq:1}
\end{equation}

When the X-ray outburst profile is compared to the OGLE IV fluxes it seems that the X-ray emission may have been triggered by what appears to be a very short duration optical flare. If it was such an optical event that produced this X-ray outburst it seems that this is a rare event compared to the typical behaviour of other Be stars. Within $\sim$14d the OGLE flux had returned to the contemporaneous base line. Though the OGLE data are somewhat sparse for this source there is no other evidence for a similar optical event over the $\sim$2000d of OGLE IV coverage. 

\color{black}

The reason for possibly associating the OGLE high point with the X-ray flare is that it is believed that the X-ray emission in \bexrb\ systems arises from accretion onto the neutron star from the Be star's circumstellar disc. The extent of the circumstellar disc is thought to be indicated by the I band flux, and so if the disc expands then material can suddenly be available for accretion onto the neutron star. This strong partnership between the I-band and the detection of X-rays is seen in many \bexrbs\ - see, for example, RX J0123.4-7321 \citep{2021coe}. In that work the authors show that when there are short-lived peaks in the I-band data there are frequently similar short-lived X-ray outbursts - see their Figure 2. So though there is only one I-band high point observed in \sw\ there is a good chance that this is showing the same effect as that seen in the data for RX J0123.4-7321 and other \bexrb systems.

\color{black}

\subsection {GAIA colour-magnitude observations}

 \color{black}
 
The optical counterpart to \sw ~ is proposed here to be of spectral type B0-0.5 IV-Ve. The intrinsic (V-I) colour of B0-0.5 IV-V spectral types is in the range -0.355 to -0.338 \citep{pm2013}. The OGLE dust maps of the LMC \citep{skowron2021} enable the precise reddening correction to be made for such an object in the LMC and that is E(V-I)=0.100. Thus the predicted observed colours of \sw ~if it were a B-type star in this range with no circumstellar disk will be (V-I) = -0.255 to -0.238. 
\cite{2010jordi} show the transformation from GAIA colour $G_{RP}$ - $G_{BP}$ to Johnson colour V-I essentially makes no numerical changes.The respective pass bands are extremely similar. 
So, from Figure~\ref{fig:gaia_col} it can be seen that the observed GAIA colours are always much redder than the expected value for a B0-0.5 IV-V star with no circumstellar disc contribution This strongly suggests the presence of a persistent disc, adding further reddening to the observed colours. 

\color{black}

\subsection{Circumstellar disc parameters}

The probable dimensions of the disk and neutron star orbit may be estimated from the results presented in this work.

The EW of the H$\alpha$ emission line may be used as a gauge to the size of the disc. It has been shown to be correlated to radii measurements from optical interferometry of nearby isolated Be stars \citep{2006ApJ...651L..53G}.
Taking the H$\alpha$ typical value as -16.8$\pm0.5$\AA ~ this predicts an H$\alpha$ emitting disk of radius (28-35) $R_\odot$ or $(2.0-2.5) \times 10^{10}$m. {\color{black} This result comes from the range of possible disk inclination angles shown in Figure 1 of \cite{2006ApJ...651L..53G} for a B2Ve star, i.e. inclinations from $0^{\circ}$ to $80^{\circ}$.}

 Also assuming, for simplicity, that the neutron star is in a circular orbit, then the period of 46.1 d permits an estimate of the orbital radius to be $1.1 \times10^{11}$ m. So, perhaps not surprisingly, the circumstellar disc size is probably much smaller than the orbit size which explains why the system is mostly in an X-ray quiescent mode. In most \bexrb systems the neutron star orbit is close to the disc size, thereby feeding material onto the neutron star and constraining further disk expansion beyond that point \citep{okazaki2001,brown2019}.

From Figure~\ref{fig:HRS_Ha} a peak separation for H$\alpha$ of $6.45 \pm 0.6$~\AA~ is determined, which implies a radius of: $r \sim 1.3 \times 10^{11} \sin^2(i)$~m \citep{Huang1972}. This would be consistent with the above RSS measurements of disc size if $i \leq 30^\circ$ (which is likely, considering that the central depression in the H$\alpha$ line is not too deep i.e. the inclination is at least not edge-on). In addition, the size of the H$\beta$ emitting region can also be estimated from the double-peak profile. From the peak separation of $5.23 \pm 0.5$~\AA ~, this results in a radius of $r \sim 1.07 \times 10^{11} \sin^2(i)$~m.

\section{Stealthy \bexrb ~systems?}

\bexrbs ~typically show a great deal of optical and X-ray variability, as witnessed by the frequent source detection rates seen by the project S-CUBED \citep{kennea2018}. This project has mapped such \bexrb ~systems in the Small Magellanic Cloud approximately weekly over several years. These variations are produced by fluctuations in the mass outflow from the Be star which drive the circumstellar disc through major size changes. Such changes in the star's behaviour are seen in the complimentary long-term optical observations of the OGLE project. As the disc changes size its interactions with the orbiting neutron star changes, and hence so does its X-ray signature. 

\sw ~is different because it exhibits a very stable Be star behaviour pattern - over at least the last 5 years - and hence presents very few opportunities for detectable X-ray emission to emerge. This picture is supported longer term by searching the archives of all X-ray telescopes that had covered this region in the past. The only previous positive detection is to be found in the ROSAT All Sky Survey : Faint Sources (rassfsc) catalogue \citep{2000voges} which lists the source as 1RXS J055007.0-681451. This catalogue contains results from observations carried out in the period July 1990 - February 1991. From that catalogue the detected signal was $(0.019\pm0.003)$ counts/s. Assuming a typical spectral power law index of -1, and the conversions provided in the catalogue, this represents a flux of $(2.2 \pm 0.3)\times 10^{-13}$ $erg/cm^{-2}s^{-1}$. Assuming the distance to the LMC of 49.97 kpc \citep{2013Piet} this then represents an X-ray luminosity of $(6.4\pm0.9)\times10^{34}$erg/s. This is $\sim$1000 times fainter than the peak luminosity reported here from the 2013 outburst. 

Furthermore, \cite{2013krimmc} analysed a deep Chandra Observatory observation of the field containing \sw ~on 2003 October. The source was not detected and the authors derived a flux limit of $\le2.7\times 10^{-15}$ $erg/cm^{-2}s^{-1}$ in the 0.3 -- 10 keV band. This represents a further factor of $\sim$100 times fainter than the ROSAT detection, and $\sim 10^{5}$ times fainter than the peak Swift luminosity.

\begin{figure}
	\includegraphics[width=8cm,angle=-00]{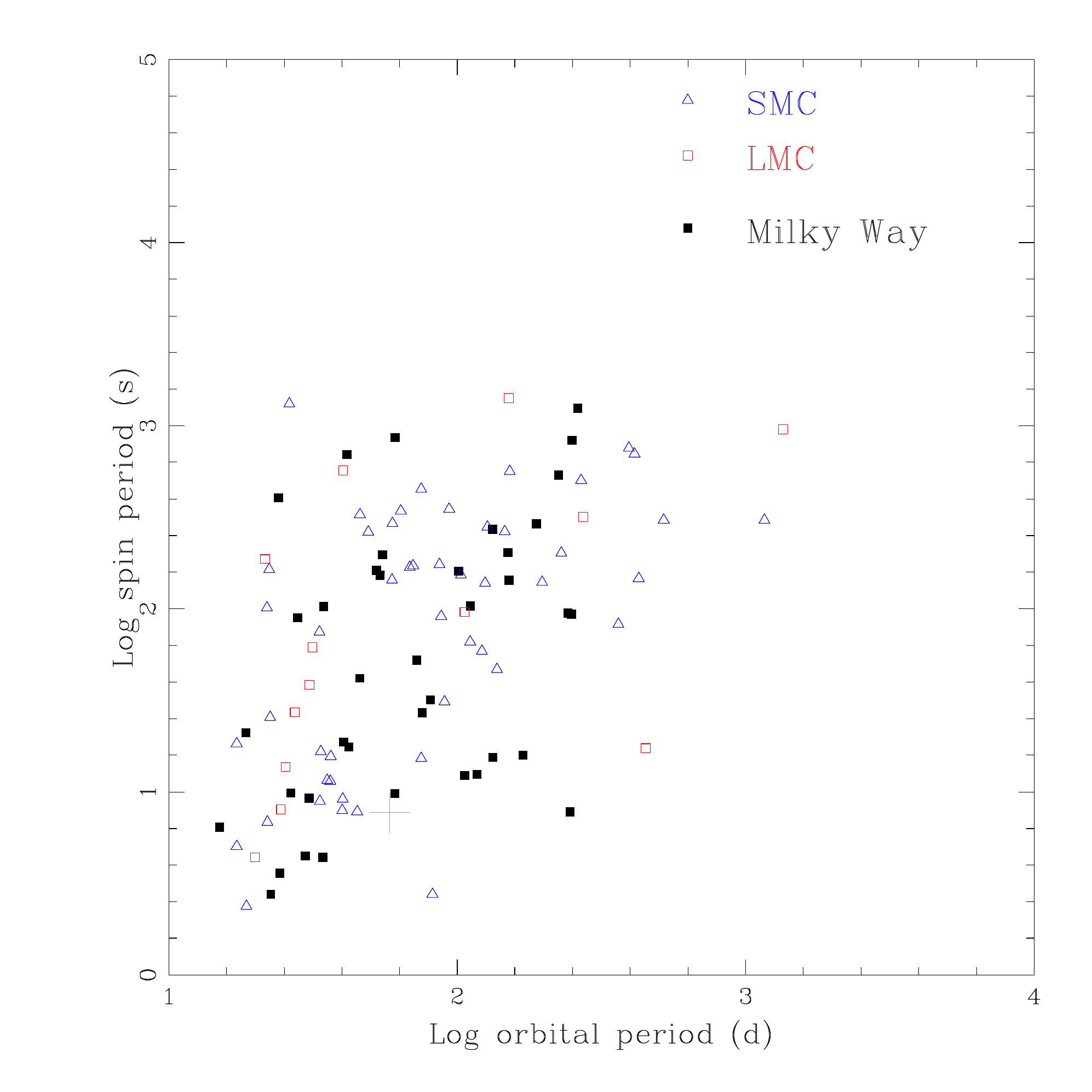}
    \caption{The "Corbet Diagram" showing the distribution of most of the known \bexrb ~systems reported in the literature. The ones in the SMC are shown as blue triangles, the LMC as red squares, and the ones in the Milky Way as black filled squares. The position of \sw ~ is shown by a green plus symbol.}
    \label{fig:corbet}
\end{figure}

The position on the well-known Corbet Diagram \citep{corbet1984} of \sw ~is worth noting - see Figure~\ref{fig:corbet}. It lies on the lower edge of the distribution with almost all other known \bexrbs with a similar orbital periods exhibiting much longer spin periods. Since these are all accretion-driven pulsar systems it is believed that their observed spin periods arise from a combination of accretion rates and magnetic field strengths (see, for example \cite{klus2014} for a review of such processes in the SMC sample of such systems). It therefore seems likely that, unless \sw ~has an unusual neutron star magnetic field strength, its position off the main distribution is a direct consequence of its very low average accretion rate.

As a result \sw ~possibly represents a class of quiescent or largely-dormant \bexrbs. Only consistent sensitive monitoring over many years will reveal how many such systems exist, and what their activity cycles are like. They could be important in the global understanding of stellar evolution and the true size of the compact object population in the Magellanic Clouds. 

\section{Conclusions}

In this work observational results have been presented on a \bexrb ~system that is rarely detected in the X-rays. Though the 2013 outburst reported here was bright (almost $10^{38}$ erg/s) it was brief, only lasting one proposed binary cycle. The short nature of this outburst, and the uncommon occurrence of such an event, is supported by the OGLE data. Careful X-ray monitoring of the Magellanic Clouds may well reveal further such stealthy systems.

\section*{Acknowledgements}
 PAE acknowledges UKSA support. JAK acknowledges support from NASA grant NAS5-00136. This work made use of data supplied by the UK Swift Science Data Centre at the University of Leicester.
 IMM and DAHB are supported by the South African National Research Foundation (NRF).

Observations reported in this paper were obtained with the Southern African Large Telescope (SALT), as part of the Large Science Programme on transients 2018-2-LSP-001 (PI: Buckley)

This work made use of data supplied by the UK Swift Science Data Centre at the
University of Leicester.

Many thanks to Poshak Gandhi and Yue Zhao for help with the GAIA data access.

This work has made use of data from the European Space Agency (ESA) mission
{\it Gaia} (\url{https://www.cosmos.esa.int/gaia}), processed by the {\it Gaia}
Data Processing and Analysis Consortium (DPAC,
\url{https://www.cosmos.esa.int/web/gaia/dpac/consortium}). Funding for the DPAC
has been provided by national institutions, in particular the institutions
participating in the {\it Gaia} Multilateral Agreement.

\section*{Data Availability}

 All X-ray data are freely available from the NASA Swift archives, and the GAIA data from the ESA archives.. The OGLE optical data in this article will be shared on any reasonable request to Andrzej Udalski of the OGLE project.



\bibliographystyle{mnras}
\bibliography{references} 



\bsp	
\label{lastpage}
\end{document}